\documentclass[12pt,notitlepage,a4paper]{article}
\pdfoutput=1

\usepackage{epsfig}
\usepackage{color,graphicx}
\usepackage{cite}
\usepackage{amssymb,amsmath}

\newcommand{\be}{\begin{equation}}
\newcommand{\ee}{\end{equation}}
\newcommand{\bea}{\begin{eqnarray}}
\newcommand{\eea}{\end{eqnarray}}

\newcommand{\lsim}{\lesssim}

\begin{document}

\begin{flushright}
CCTP-2011-01 
\end{flushright}

\begin{center}  

\vskip 2cm 

\centerline{\Large {\bf A holographic quantum Hall model at integer filling}}
\vskip 1cm

\renewcommand{\thefootnote}{\fnsymbol{footnote}}

\centerline{Niko Jokela,${}^{1,2}$\footnote{najokela@physics.technion.ac.il}
Matti J\"arvinen,${}^3$\footnote{mjarvine@physics.uoc.gr} 
and Matthew Lippert${}^3$\footnote{mlippert@physics.uoc.gr}}

\vskip .5cm
${}^1${\small \sl Department of Physics} \\
{\small \sl Technion, Haifa 32000, Israel} 

\vskip 0.2cm
${}^2${\small \sl Department of Mathematics and Physics} \\
{\small \sl University of Haifa at Oranim, Tivon 36006, Israel} 

\vskip 0.2cm
${}^3${\small \sl Crete Center for Theoretical Physics} \\
{\small \sl Department of Physics} \\
{\small \sl University of Crete,  71003 Heraklion, Greece}

\end{center}

\vskip 0.3 cm

\setcounter{footnote}{0}
\renewcommand{\thefootnote}{\arabic{footnote}}

\begin{abstract}

\noindent We construct a holographic model of a system of strongly-coupled fermions in 2+1 dimensions based on a D8-brane probe in the background of D2-branes.  The Minkowski embeddings of the D8-brane represent gapped quantum Hall states with filling fraction one.  By computing the conductivity and phase structure, we find results qualitatively similar to the experimental observations and also to the recent D3-D7' model.
\end{abstract}

\newpage

\section{Introduction}
The quantum Hall effect (QHE) describes the generic behavior of gapped fermions in 2+1 dimensions with broken parity.\footnote{For a review see, for example, \cite{girvin}.}   A typical physical manifestation involves a low-temperature electron gas confined to the interface of a heterojunction in a strong, transverse magnetic field.  As a result of its striking and robust experimental signatures, the QHE has been a vigorous area of research since its discovery three decades ago.

Much of the recent focus has been on the fractional quantum Hall effect (FQHE), in which the filling fraction $\nu$, the ratio of the charge density to the magnetic field in units of $e/h$, takes certain fractional values.  One reason a full explanation of the FQHE remains elusive is that it results from strongly-coupled dynamics, as evidenced by interesting features such as fractionally-charged quasiparticles \cite{quasiparticles1, quasiparticles2}.  

The integer quantum Hall effect (IQHE), where $\nu \in \mathbb Z$, was discovered first because it is much easier to detect experimentally.  It is also much better understood since, for the most part, it can be understood in terms of free quasiparticles organized into Landau levels.  However, recent interferometry experiments \cite{visibility} indicate that perhaps the situation is not so simple and interactions are indeed important. 

Gauge/gravity duality has proven to be a powerful tool for investigating strongly-coupled systems, such as those arising in many interesting condensed matter systems. The QHE, in particular, has proven to be amenable to holographic study.  For example, phenomenological ``bottom-up'' models \cite{KeskiVakkuri:2008eb, Kachru1, Kachru2, Bayntun:2010nx, Gubankova:2010rc} of dyonic black AdS black holes have been proposed to capture certain observed features of QHE physics, in particular those related to $SL(2,\mathbb Z)$ duality, but do not include a mass gap.

Brane constructions realizing quantum Hall physics \cite{Brodie:2000yz, Bergman:2001qg, Hellerman:2001yv} predate the current holographic approach.  In more recent brane models \cite{Rey:2008zz, Davis:2008nv, 
 Myers:2008me, Fujita:2009kw, Hikida:2009tp, Alanen:2009cn, Hong:2010sb, Bergman:2010gm, Belhaj:2010iw}, 2+1-dimensional brane intersections with $\# ND=6$ yield a low-energy spectrum of fundamental fermions.\footnote{See also \cite{Fujita:2010pj} for a M5-brane model of the 1+1-dimensional edge of a QH system.}  These models are related to the Sakai-Sugimoto model of QCD in 3+1 dimensions \cite{SS}.

The philosophy of this type of ``top-down'' approach is essentially pragmatic.  Starting with a tractable string theory solution yields a holographic dual of the known field theory whose properties can be studied and which can hopefully then be matched to interesting physical systems.  Previous models have been based in different ways on D3-D7 intersections, but there are other possibilities which should be analyzed in order to determine which features are robust and universal across models and the degree to which the others vary.

With this approach in mind, we construct a holographic model of $N$ species of 2+1-dimensional fermions coupled to a strongly-interacting 2+1-dimensional gauge field using a probe D8-brane nontrivially embedded in a thermal D2-brane background.  We refer to this model as the D2-D8' model with the prime denoting the nonstandard embedding and emphasizing the analogy with the D3-D7' model of \cite{Bergman:2010gm}.   The embedding is stabilized by a worldvolume flux wrapping an internal cycle.  The charge density and transverse magnetic field are holographically encoded in additional components of the worldvolume gauge field.

We find that the two classes of embeddings correspond to different phases.  Minkowski embeddings (MN), for which the D8-brane smoothly caps off before entering the black hole horizon, represent gapped QH states.   The dynamically-induced mass gap for charged excitations is roughly given by the distance between the tip of the D8-brane and the horizon.  Black hole embeddings (BH), where the D8-brane crosses the horizon, are ungapped, metallic states.  For a given charge density, there is a single MN embedding; this solution has filling fraction per fermion species $\nu = 1$, meaning that the D2-D8' system is a holographic model of the IQHE. 

We compute the electrical conductivity and the phase diagram and find some qualitative agreement with experimental observations.  As expected in the QH state, the Hall conductivity $\sigma_{xy} = \frac{\nu}{2\pi}$ and the longitudinal conductivity $\sigma_{xx}=0$.  The QH state exists only at low temperature; we find a first-order phase transition rather than a crossover to a metallic state when the temperature is as large as the mass gap.  
Moving away from the QH state by changing the charge density or the magnetic field again leads to a phase transition to the metallic state rather than a crossover.  The conductivity of the metallic state qualitatively obeys the semicircle law \cite{Dykhne, Hilke1, Hilke2, Bayntun:2010nx}, albeit in a very rough sense, tracing out something like a hyperbola in the $(\sigma_{xy},\sigma_{xx})$-plane as the magnetic field is varied.

Some discrepancies with real QH systems are due to unphysical simplifications of the model.  In particular, the lack of impurities, which are required for the characteristic plateaux in the Hall conductivity, is the reason that the MN embedding only exists, for a given charge density, at a single value of the magnetic field.  In addition, the conductivity of the perfectly clean metallic phase has unphysical behavior as zero temperature is approached.

In many respects, this model reproduces many of the features found in the D3-D7' model of \cite{Bergman:2010gm}.  One important difference is that while the D2-D8' model is dual to a fully 2+1-dimensional boundary theory, the D3-D7' describes a defect theory in which fermions are confined to 2+1 dimensions while the gauge bosons propagate in 3+1 dimensions.  Another significant difference is that rather than exhibiting the IQHE, the D3-D7' models the FQHE with a filling fraction set by the internal fluxes.

In the following section, we describe in detail the D2-D8' system, and in Section \ref{sec:embeddings}, we find the D8-brane embeddings which correspond to the quantum Hall and metallic states.  In Section  \ref{sec:conductivity} we compute the conductivity of these two states, and in Section \ref{sec:phasestructure} we analyze the phase structure of the model.  We conclude with a summary and discussion in Section \ref{sec:discussion}.

\section{The D2-D8' system}\label{sec:setup}
In order to construct a nonsupersymmetric system of fermions in 2+1 dimensions, we consider a setup of $N$ D2-branes and a single D8-brane with $\# ND=6$.  Internal flux is needed to stabilize this nonsupersymmetric system.  The low-energy spectrum of bifundamental strings in a $\# ND=6$ system contains only charged fermions and no charged bosons.  In this case, the boundary field theory is 2+1-dimensional SYM coupled to $N$ species of charged fermions.  Similar constructions have been used, for example, in the D3-D7 \cite{Myers:2008me} and D3-D7' \cite{Bergman:2010gm} models.  

We will work in the limit $N \gg 1$, so we can treat the D8-brane as probes in the background sourced by the D2-branes.  We will also take the usual 't Hooft limit, where $g_s N \gg 1$.

\subsection{D2-brane background}
We consider the near-horizon background of $N$ thermal D2-branes.\footnote{See \cite{Itzhaki:1998dd} for a thorough discussion of the D2-brane background and the regimes of validity of various descriptions.}  The 10-dimensional metric is 
\be
L^{-2} ds^2_{10} = u^\frac{5}{2}\left(-f(u)dt^2+dx^2+dy^2\right)+u^{-\frac{5}{2}}\left(\frac{du^2}{f(u)}+ u^2 d\Omega_{S^6}^2\right) \ ,
\ee
where $L^5 = 6\pi^2 g_s N l_s^5$ and the thermal factor $f(u)=1-\left(\frac{u_T}{u}\right)^5$.  Note that this space is not asymptotically AdS.  The corresponding Hawking temperature is $T= \frac{5}{4\pi L} u_T^{3/2}$.  Furthermore, the background dilaton and the RR 3-form potential are
\be 
e^\phi = g_s\left(\frac{L}{U}\right)^\frac{5}{4}\quad ,\quad C^{(3)} = -u^5 L^3 dt\wedge dx\wedge dy \ .
\ee

We will work in  dimensionless (lowercase) coordinates, which are related to the physical dimensionful (uppercase) coordinates as: $x^\mu = X^\mu/L$, $u = U/L$.  In general, lowercase Latin letters will denote dimensionless quantities, and uppercase letters will denote the corresponding physical quantities.

We choose to parametrize the $S^6$ in a somewhat unconventional way, analogous to the D3-D5 system of \cite{Myers:2008me} and the D3-D7' model of \cite{Bergman:2010gm}, as an $S^2 \times S^3$ fibered over an 
interval.  Introduce the angles $\psi$, $\theta_1$, $\phi_1$, $\theta_2$, $\phi_2$, and $\xi$ such that:
\be
d\Omega_{S^6}^2 = d\psi^2 + \sin^2\psi \left(d\theta_1^2 + \sin^2\theta_1 d\phi_1^2\right) + \cos^2\psi \left(d\xi^2 + \sin^2\xi d\theta_2^2+ \sin^2\xi \sin^2\theta_2 d\phi_2^2\right) \ .
\ee
The angle $\psi$ ranges from 0 to $\pi/2$, the angles $\xi$, $\theta_1$, and $\theta_2$ range from 0 to $\pi$, and the angles $\phi_1$ and $\phi_2$ range from 0 to $2\pi$.

We will need the RR 5-form potential $C^{(5)}$ dual to $C^{(3)}$.  Taking the Poincar\'e dual of the 4-form field strength $G^{(4)} = dC^{(3)}$ gives
\be
G^{(6)} = L^5 \sin^2\psi \cos^3\psi \, d\psi\wedge d\Omega_{S^2} \wedge d\Omega_{S^3} \ .
\ee
This can be integrated with respect to $\psi$ to find
\be
C^{(5)} = c(\psi) L^5  d\Omega_{S^2} \wedge d\Omega_{S^3} \ ,
\ee
where $c(\psi) = \frac{5}{8}\left(\sin\psi - \frac{1}{6}\sin(3\psi)- \frac{1}{10}\sin(5\psi)\right)$.  We have fixed the constant of integration by choosing the gauge $c(0) = 0$.

\subsection{D8-brane probe}
The probe D8-brane fills the spacetime directions $t$, $x$, $y$, and the radial direction $u$ and wraps the $S^2$ and the $S^3$.  The embedding in the $\psi$-direction will be a function of $u$, as $\psi(u)$.  The induced metric reads 
\bea
L^{-2}ds^2_{D8} &=& u^\frac{5}{2}\left(-f dt^2+dx^2+dy^2\right)+u^{-\frac{5}{2}}\left(\frac{1}{f} +u^2\psi'^2\right)du^2 \nonumber\\
&&+ u^{-\frac{1}{2}}\sin^2\psi d\Omega_2^2 + u^{-\frac{1}{2}}\cos^2\psi d\Omega_{3}^2 \ .
\eea

We would like to consider a boundary system at nonzero charge density in a background magnetic field.  Although the boundary U(1) field is non-dynamical and the corresponding current is global, we can still mimic the effects of a background field on a local charge current.\footnote{There is an extensive literature on this issue; for example, see \cite{Marolf:2006nd, Domenech:2010nf, Maeda:2010br}.}  Holographically, this is accomplished by turning on components of the worldvolume gauge field; specifically, a constant field strength in the spatial directions corresponds to a transverse magnetic field, and nonzero charge density is dual to a radially varying 
electric field:\footnote{We choose a gauge where $a_u = 0$.}
\bea
2\pi\alpha' F_{xy} &=& h \\
2\pi\alpha' F_{u0} &=& a_0'(u) \ .
\eea
In addition, as we will see, in order to stabilize the embedding, we will also turn on a magnetic field wrapping the internal $S^2$:
\be
2\pi\alpha' F = b L^2  \, d\Omega_2 \ .
\ee
Turning on this wrapped flux can be alternatively viewed as dissolving smeared D6-branes into the D8-brane.  

The DBI action for the D8-brane is then
\bea
 S_{DBI} &=&  - \mu_8 \int d^9x \, e^{-\phi} \sqrt{-\det(g_{\mu\nu} + 2\pi\alpha' F_{\mu\nu})}  \\
&=&  - \mathcal N \int du  \, u^\frac{5}{2} \cos^3\psi \sqrt{\left(1+f u^2\psi'^2\right)\left(b^2u + \sin^4\psi\right)\left(1+\frac{h^2}{u^5}\right)} \quad ,
\eea
where $\mathcal N = 8 \pi^3 T_8 v_3 L^9$ and $v_3$ is the dimensionless volume of spacetime.  As a result of the charge density and the magnetic field, the Chern-Simons action has a nonzero component:
\be\label{CSaction}
S_{CS} =- \frac{T_8}{2}(2\pi\alpha')^2\int C^{(5)}\wedge F\wedge F = \mathcal N \int du \, c(\psi) ha_0' \ .
\ee
In order to make the Chern-Simons action invariant under gauge transformations of $C^{(5)}$, we also need to add the following boundary term \cite{Bergman:2008qv, Bergman:2010gm}, which affects the energy but not the equations of motion:
\be
\label{boundaryaction}
S_{bdry} = - \frac{T_8}{2}(2\pi\alpha')^2 \int d\left(C^{(5)}\wedge F\wedge A\right) =  - \mathcal N c(\psi) ha_0 |^\infty_{u_{min}} \ .
\ee
Since we have chosen the gauge $c(\psi=0) = 0$, there is no contribution at $u=\infty$ and so $S_{bdry} = \mathcal N c(u_{min}) ha_0(u_{min})$.

From this action, we compute the equation of motion for $\psi$:
\bea
\label{psieoma0}
\partial_u\left(u^2 g(u) \left(1 + \frac{h^2}{u^5}\right) \psi'\right) &=& \frac{f u^5}{g(u)}\cos^5\psi\sin\psi\left(2\cos^2\psi\sin^2\psi - 3\left(b^2u + \sin^4\psi\right)\right) \nonumber\\
&& - \partial_\psi c(\psi)ha_0' \ ,
\eea
where we have defined
\be
\label{ga0}
g(u) = \frac{f u^{5/2}\cos^3\psi \sqrt{b^2u + \sin^4\psi}}{\sqrt{\left(1+f u^2\psi'^2-a'^2_0\right)\left(1 + \frac{h^2}{u^5}\right)}} \ .
\ee
The action is independent of $a_0$, so the equation of motion for $a_0$ can be integrated once to give
\be
\frac{g}{f}\left(1 + \frac{h^2}{u^5}\right) a_0' = d-hc(\psi) \equiv \tilde d(u) \ ,
\ee
where $d$ is the constant of integration and $\tilde d(u)$ is the radial displacement field.  In terms of the boundary theory, $d$ is the total charge density, while $\tilde d(u)$ is the charge due to sources located in the bulk at radial positions below $u$.

In order to decouple the equations and express the $\psi$ equation in terms of the charge, we can solve for $a_0'$ in terms of $\tilde d$, obtaining
\be
a_0' = \tilde d \sqrt{\frac{1+fu^2\psi'^2}{\tilde d^2 + u^5\cos^6\psi \left(b^2u + \sin^4\psi\right)\left(1 + \frac{h^2}{u^5}\right) }} \quad .
\ee
We can now write (\ref{ga0}) as
\be
\label{g}
g = \frac{f}{1+\frac{h^2}{u^5}} \sqrt{\frac{\tilde d^2 + u^5\cos^6\psi\left(1 + \frac{h^2}{u^5}\right)\left(b^2u + \sin^4\psi\right)}{1+f u^2\psi'^2}} \quad ,
\ee
and (\ref{psieoma0}) can be rewritten as
\bea
\label{psieom}
\partial_u\left(u^2 g \left(1 + \frac{h^2}{u^5}\right) \psi'\right) &=& \frac{f u^5}{g} \cos^5\psi\sin\psi\left(2\cos^2\psi\sin^2\psi - 3\left(b^2u + \sin^4\psi\right)\right) \nonumber \\ &&- 5 \frac{f}{g}\frac{h \tilde d}{1+\frac{h^2}{u^5}}\cos^3\psi\sin^2\psi \ .
\eea

\subsection{UV asymptotics}
Because supersymmetry is completely broken in this system, one might expect that the D8-brane embedding would be unstable.  In flat space, a D2-brane and a D8-brane which are parallel with $\# ND=6$ repel each other, leading one to anticipate a similar instability after taking the near-horizon limit \cite{Rey:2008zz}.   Rather than analyzing the fluctuation spectrum directly, we will take the simpler approach of computing the dimension of the operator dual to $\psi$ to see if it has a nonzero imaginary part.  The details of the calculations are presented in Appendix \ref{UVasymptotics}.  

As expected, we see a tachyonic mode in the D8-brane embedding when there is no flux wrapping the internal $S^2$; the D8-brane slips to one of the poles of the $S^6$.  However, as in other similar systems, such as the D3-D7 model of \cite{Myers:2008me} and D3-D7' model of \cite{Bergman:2010gm}, worldvolume fluxes can stabilize the probe brane.  For any nonzero $b$, there are apparently stable D8-brane embeddings.  Of course, a detailed analysis of the fluctuations will be required to guarantee stability, but at least the most dangerous tachyon has been removed.  

This result is in contrast to what has been seen in the analogous D3-D7 systems \cite{Myers:2008me, Bergman:2010gm}, where a minimum nonzero flux was required.\footnote{Note, however, that in the D3-D7' system of \cite{Bergman:2010gm}, for the special cases of $\psi_\infty = 0$ and $\pi/2$, there is no lower bound on the flux required for stability.} Also unlike the D3-D7'  system, both the asymptotic angle, $\psi_\infty = 0$, and the subleading behavior, $\Delta = -1$ or $-3$, are fixed and are independent of the size of the flux.  

In the flat space embedding, the fermion mass is given by the minimal distance between the D2-brane and the D8-brane.  As in the D3-D7' system, in the near-horizon limit, we identify the coefficient of the leading term in the asymptotic expansion of $\psi$ with the mass and the coefficient of the subleading term as the chiral condensate: 
\be
\psi = \frac{m}{u} - \frac{c}{u^3} + \mathcal O(u^{-4}) \ .
\ee


\section{Embeddings}
\label{sec:embeddings}

Having understood the UV limit of the D8-brane embedding, we now consider the full embedding.  There are two classes with different IR behavior, black hole (BH) embeddings, which enter the horizon at $u_T$, and Minkowski (MN) embeddings, which smoothly cap off at some $u_0 > u_T$.  As argued in \cite{Bergman:2010gm}, the MN embeddings holographically reproduce the properties of a quantum Hall fluid, while the BH solutions exhibit the metallic behavior characteristic of the regions between quantum Hall plateaux.

For the figures presented here, we have chosen values of the parameters $u_T$, $m$, $d$, $h$, and $b$ for illustrative purposes.   The scaling $u \to \lambda u$, $u_T \to \lambda u_T$, $m \to \lambda m$, $d \to \lambda^{5/2} d$, $h \to \lambda^{5/2} h$, and $b \to \lambda^{-1/2} b$ is a symmetry of the equations of motion, and so there is, in fact, only a four-dimensional parameter space.  We find that qualitative results are independent of these choices.  However, we can not conclusively rule out the possibility of different behavior in particular corners of parameter space.

\subsection{MN embeddings}\label{sec:MNembeddings}

For the D8-brane to smoothly end at $u_0$, one of the internal spheres must shrink to zero at the tip.  Because of the flux wrapped on the $S^2$, we choose the $S^3$ to shrink instead, giving $\psi(u_0) = \pi/2$ and as $u \to u_0$, $\psi' \to  -\infty$.  As in the D3-D7' model \cite{Bergman:2010gm}, because the tip of the D8-brane cannot support a worldvolume charge, all the charge must be induced by the magnetic field via the Chern-Simons term (\ref{CSaction}).  This implies that for MN embeddings, the ratio of the holographic charge density $d$ and the magnetic field $h$ is fixed:
\be
\frac{d}{h} = c(\pi/2) = \frac{2}{3} \ .
\ee
The filling fraction per fermion species $\nu$ is defined by the ratio of the physical charge density $D = \frac{(2\pi\alpha')\mathcal N}{L^4 v_3} d$ to the physical magnetic field $H = \frac{h}{2\pi\alpha'}$ divided by the number of species $N$.  We find that
\be
\nu = \frac{2\pi}{N} \frac{D}{H} = \frac{3}{2} \frac{d}{h} = 1 \ .
\ee
The MN embedding is therefore a holographic model of an integer, rather than fractional, quantum Hall fluid. 

Alternately, we can compute $\nu$ from the 3-dimensional effective Chern-Simons term
\be
\label{effectiveCS}
S_{CS} = -\frac{1}{2}T_8 (2\pi\alpha')^2 \int G^{(6)}\wedge A\wedge F \ .
\ee
Because our MN embedding goes from $\psi = 0$ to $\pi/2$, we integrate over the whole $S^6$ to find
\be
 \int G^{(6)} = 8\pi^3 L^5 \int_{0}^{\pi/2} d\psi \, c \hspace{0.2mm} '(\psi) = (2\pi l_s)^5 g_s N \ .
\ee
Plugging this in to (\ref{effectiveCS}) gives:
\be
S_{CS} = -\frac{N}{4\pi}\int A\wedge F \ .
\ee
The level $k$ can be read off from the coefficient of the Chern-Simons term $\frac{k}{4\pi}$; we see directly that $k = N$.  And since the level $k$ is also given by the total filling fraction $N\nu$, this gives the same result as above.

\begin{figure}[ht]
\center
\includegraphics[width=0.65\textwidth]{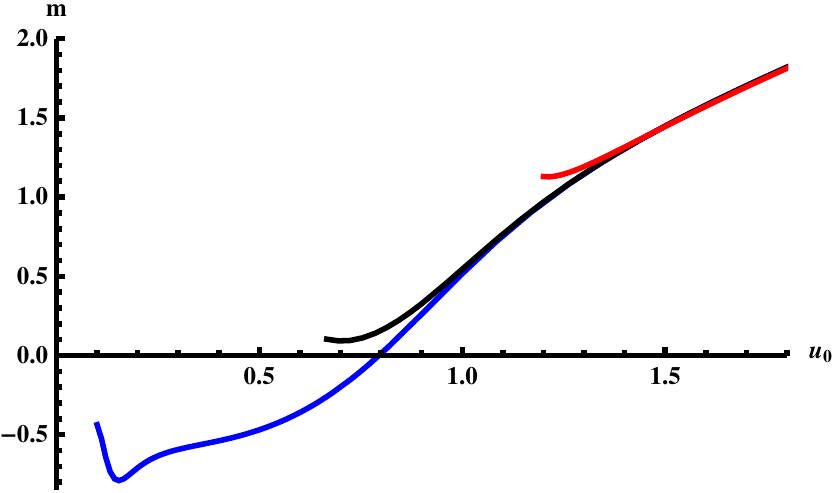} 
\caption{
The mass $m$ as a function of the position of the D8-brane tip $u_0$, for $b=1$, $d=1$, and $h=1.5$.  The bottom, blue curve shows $u_T = 0.1$, the middle, black $u_T = 0.666$, and the top, red $u_T = 1.2$.  Notice that the minimum mass $m_{min}$ increases with temperature.  For $m=0.1$, at the low temperature there is one MN solution, two solutions at the intermediate temperature, and then no solutions at the high temperature.}
\label{fig:mvsu0-uT}
\end{figure}

To find the complete embedding, we solve equation (\ref{psieom}) numerically by shooting out from the IR tip and extracting $m$ and $c$ from the UV asymptotics.  

For certain initial conditions, $\psi(u)$ becomes negative at some $u$ and then asymptotically approaches $\psi = 0$ from below, resulting in $m < 0$.  While there is nothing a priori wrong with a negative fermion mass, in this case it is a sign that the low-energy approximation is breaking down.  When, at some finite $u$, $\psi(u)$ goes to zero with finite $\psi'$, the D8-brane self-intersects at a conical singularity.  For such a configuration, massive modes of the D8-brane, which are excluded from the DBI action, become light and, in fact, tachyonic, and the embedding is unstable.  As a result, we only consider positive masses.

We show the dependence of $m$ on $u_0$ for several temperatures in Fig.~\ref{fig:mvsu0-uT}.   For a given temperature, and fixed $b$ and $d$, there is a minimum mass $m_{min}$ for which a solution exists.  At temperatures sufficiently below the mass gap, \emph{i.e.}, $u_T \ll u_0$, this minimum mass is negative, and there exists a unique solution for any given positive $m$.  However, $m_{min}$ increases with the temperature, becoming positive, and eventually, when $m_{min} \sim m$,  a second MN embedding at smaller $u_0$ appears.  As we shall see in Section \ref{sec:phasestructure}, this embedding is thermodynamically unstable.  We also expect it to be perturbatively unstable as well.  As the temperature is increased further, the two solutions merge and disappear; at sufficiently high temperature, there are no MN embeddings with mass $m$ because $m_{min} > m$.

\begin{figure}[!ht]
\center
\includegraphics[width=0.65\textwidth]{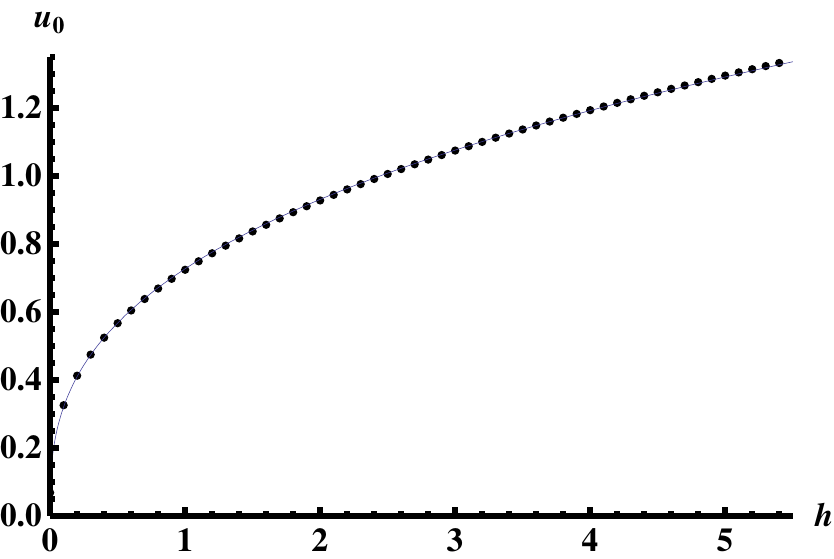} 
\caption{The mass gap in the MN embedding, $m_{gap} = u_0$, as a function of the magnetic field $h$ for $u_T= 0.01$, $b=1$, and $m =0.1$.  The blue curve is a fit to $u_0 \propto h^{0.36}$.}
\label{fig:massgap}
\end{figure}

A characteristic property of a quantum Hall fluid is a mass gap, and, as shown in \cite{Bergman:2010gm, Jokela:2010nu}, MN embeddings holographically realize the mass gap, denoted by $m_{gap}$, by avoiding the horizon.  Although we postpone a detailed analysis of the fluctuation spectrum for a future work, we can easily estimate the mass gap for charged excitations.  Charged quasiparticles are holographically dual to strings stretched from the horizon to the D8-brane tip, and at zero temperature, the mass of such strings, and thus the mass gap for charged excitations, is simply $m_{gap} = u_0$.  For an integer quantum Hall fluid, the mass gap increases linearly with the magnetic field; here, however, the gap scales as $h^{0.36}$ to very good approximation, as shown in Fig.~\ref{fig:massgap}.

\subsection{BH embeddings}
\label{sec:BHembeddings}

Black hole embeddings are those which enter the horizon.  As with MN embeddings, we solve (\ref{psieom}) numerically by shooting, in this case from the horizon.  We need only to choose a single IR boundary condition, the location where the D8-brane intersects the horizon $\psi_T$, because equation (\ref{psieom}) becomes first-order in the near-horizon limit.  Some initial conditions lead to embeddings with negative mass, which, as discussed in Section \ref{sec:MNembeddings}, we disregard as unstable artifacts.

\begin{figure}[ht]
\center
\includegraphics[width=0.65\textwidth]{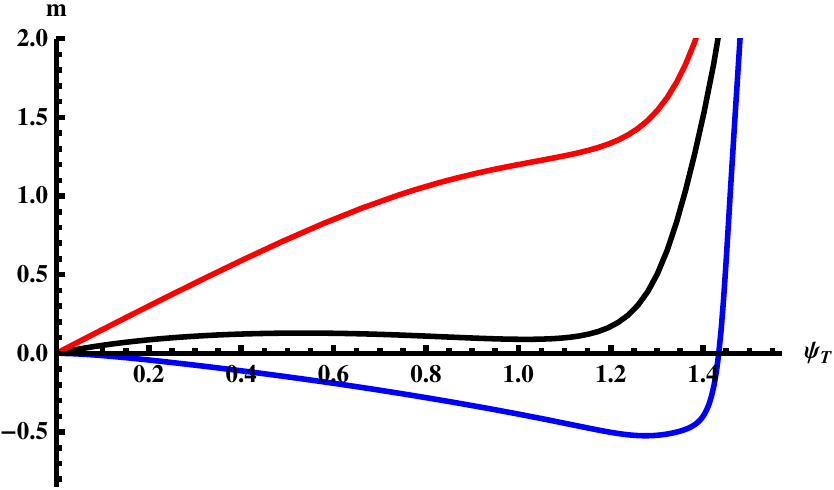} 
\caption{The mass $m$ as a function of $\psi_T$, with $b=1$, $d = 1$, and $h=1.25$.  Curves for three temperatures are plotted: bottom, blue for $u_T = 0.1$, middle, black for $u_T = 0.53$, and top, red for $u_T = 1.2$.  For $m=0.1$, for example, at the low temperature there is a single solution at $\psi_T \approx 1.44$.  There are three solutions $\psi_T \approx 0.25$, $0.88$, and $1.10$ at the intermediate temperature, and at the high temperature, one solution at $\psi_T \approx 0.07$.}
\label{fig:psiTvsrT}
\end{figure}

\begin{figure}[!b]
\center
\includegraphics[width=0.46\textwidth]{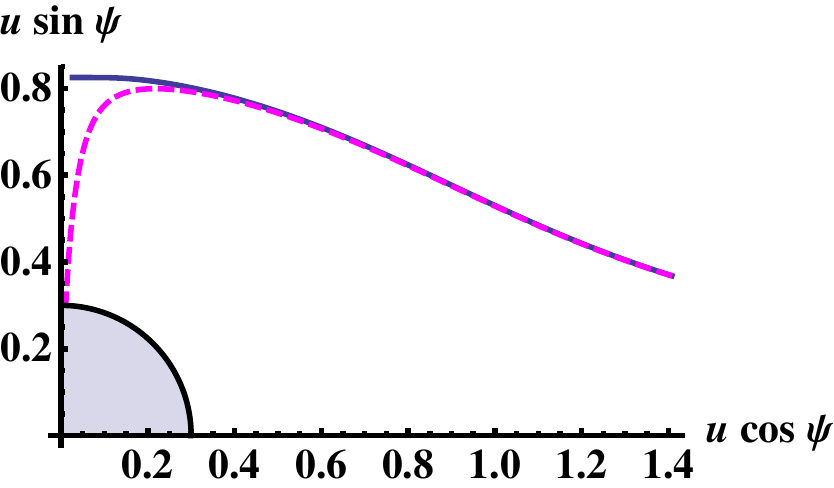} 
\hspace{0.5cm}
\includegraphics[width=0.46\textwidth]{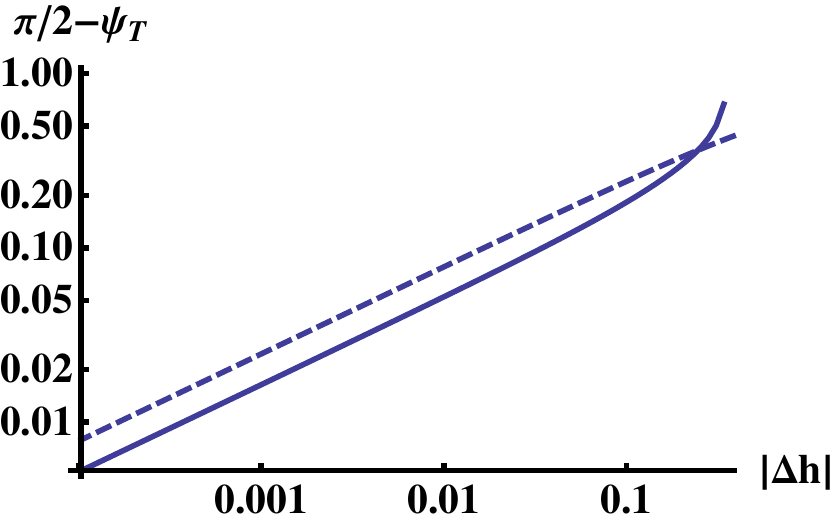} 
\caption{In the left panel, as the magnetic field approaches locking, $h \to \frac{3}{2} d$, the large $\psi_T$ BH embedding (dashed pink line) becomes increasingly spiky, $\psi_T \sim \pi/2$, and approaches the MN embedding (solid blue line).  Both embeddings are for $m = 0.1$, $b=d=1$, and $u_T=0.3$.  The MN embedding has $h = 1.5$, while the BH embedding is $h = 1.49$.  The black hole is indicated by the quarter circle near the origin.  In the right panel, the approach to a MN embedding, also for $m=0.1$, $b=d=1$, is shown as a log-log plot of $\pi/2 - \psi_T$ versus $|\Delta h|$, illustrating $\frac{\pi}{2} - \psi_T \sim |\Delta h|^{1/2}$.  The solid line is for $\Delta h > 0$, and the dashed line for $\Delta h < 0$.}
\label{fig:spike}
\end{figure}

For generic $h$ and $d$, the behavior of $m$ as a function of $\psi_T$ is as shown in Fig.~\ref{fig:psiTvsrT}.  At low temperature, there is a single solution at $\psi_T \lsim \pi/2$ for any given positive mass.  But, as the 
temperature is raised, two embeddings appear; as we will see in Section \ref{sec:phasestructure}, the one at smaller $\psi_T$ is stable, while the other is not.  At still higher temperature, the unstable solution merges with the large $\psi_T$ solution, and they disappear, leaving only the small $\psi_T$ embedding.

As in \cite{Bergman:2010gm}, as the magnetic field is tuned such that $\nu \to 1$, the large $\psi_T$ BH solution starts to resemble a MN embedding connected to the horizon by a thin spike.  Fig.~\ref{fig:spike} shows how $\psi_T \to \pi/2$ as the magnetic field is tuned to $\nu = 1$.  We find that for small $\Delta h \equiv h - \frac{3}{2} d$,
\be
\frac{\pi}{2} - \psi_T \sim |\Delta h|^{1/2} \ . 
\ee

Although the transition from a BH to a MN embedding at $\Delta h = 0$ appears discontinuous, certain physical quantities, such as the conductivities are continuous, as we will see in Section \ref{sec:conductivity}.  We will investigate more closely in Section \ref{sec:transitionfromlocking} and find a phase structure analogous to the MN to BH transition in the supersymmetric D3-D7 system \cite{Ghoroku:2007re, Mateos:2007vc}.


\section{Conductivity}
\label{sec:conductivity}
To compute the conductivities in both the BH and MN states, we follow the well-known Karch-O'Bannon procedure \cite{Karch:2007pd,O'Bannon:2007in}.  We holographically introduce an electric field in the $x$-direction and currents in the $x$- and $y$-directions, as follows:
\bea
2\pi\alpha'F_{0x} &=& e \\
2\pi\alpha'F_{ux} &=& a'_x \\
2\pi\alpha'F_{uy} &=& a'_y \ .
\eea

The DBI action is now
\be
S_{DBI} = -\mathcal N \int du  \, u^\frac{5}{2} \cos^3\psi \sqrt{ Y \left(b^2u + \sin^4\psi\right)} \ ,
\ee
where
\bea
Y &=& \left(1+\frac{h^2}{u^5}-\frac{e^2}{fu^5}\right)\left(1+f u^2\psi'^2\right)-\left(1+\frac{h^2}{u^5}\right)a_0'^2 \nonumber\\
&&+f a_x'^2+f\left(1-\frac{e^2}{fu^5}\right)a_y'^2-2a_0'a_y' \frac{eh}{u^5} \ .
\eea
There is also an additional term in the Chern-Simons action
\be
S_{CS} = \mathcal N \int du \, c(\psi) \left(ha_0' + ea_y'\right) \quad .
\ee

The equations of motion for the gauge fields $a_\mu$ can be computed and integrated once, introducing constants of integration $d$, $j_x$, and $j_y$ as follows:
\bea
\tilde d \equiv d-hc(\psi) &=& \frac{g}{f}\left( \left(1+\frac{h^2}{u^5}\right)a_0' + \frac{eh}{u^5}a_y'\right) \\
 j_x &=& g a_x' \\
\tilde j_y \equiv j_y - ec(\psi) &=& g\left(\left(1-\frac{e^2}{f u^5}\right)a_y' - \frac{eh}{f u^5}a_0'\right) \ ,
\eea
where the function $g$ is now
\be
g = f u^{5/2}\cos^3\psi \sqrt{\frac{b^2u + \sin^4\psi}{Y}} \quad .
\ee



Using some algebra, we can solve for the gauge fields in terms of the currents:
\bea
a_0' &=& \left(f\left(1-\frac{e^2}{f u^5}\right)\tilde d - \frac{eh}{u^5}\tilde j_y\right) \sqrt{\frac{1+f u^2\psi'^2}{X}} \\
a_x' &=& \left(1+\frac{h^2}{u^5}-\frac{e^2}{f u^5}\right) j_x \sqrt{\frac{1+f u^2\psi'^2}{X}} \\
a_y' &=& \left(\left(1+\frac{h^2}{fu^5}\right)\tilde j_y - \frac{eh}{u^5}\tilde d\right) \sqrt{\frac{1+f u^2\psi'^2}{X}} \ ,
\eea
where
\bea
X &=& f  \left(1+\frac{h^2}{u^5}-\frac{e^2}{f u^5}\right) \left(f u^5\cos^6\psi\left(b^2u+\sin^4\psi\right) + f \tilde d^2 - j_x^2 - \tilde j_y^2\right) \nonumber \\
&& - \frac{\left(f h\tilde d + e\tilde j_y\right)^2}{u^5} \quad .
\eea

\subsection{Quantum Hall fluid}
\label{sec:QHconductivity}

For the MN embeddings, we use a modified version of the Karch-O'Bannon method \cite{Bergman:2010gm}; the conductivities can be computed from the requirement that the gauge fields be regular at the tip $u_0$, which is accomplished by 
\be
\tilde d = j_x = \tilde j_y = 0 \quad .
\ee
The vanishing of $\tilde d(u_0)$ implies
\be
\frac{d}{h} = c(\pi/2) = 2/3 \quad .
\ee
Because $j_x = 0$, there is no longitudinal current and 
\be
\label{sigmaxxQH}
 \sigma_{xx} = 0 \ .  
\ee
Lastly, the vanishing of $\tilde j_y$ gives $j_y = e c(\pi/2)$, from which we find the physical Hall conductivity
\be
\sigma_{xy} \equiv \frac{J_y}{E} = \frac{(2\pi\alpha')^2\mathcal N}{L^4 V_3} \frac{j_y}{e} = \frac{N}{2\pi} \ . 
\ee
Previously, in Section \ref{sec:MNembeddings}, we found the filling fraction per fermion species $\nu = 1$.  Therefore, we find
\be
\label{sigmaxyQH}
\sigma_{xy} = \frac{N\nu}{2\pi}
\ee
which is the expected transverse conductivity in a quantum Hall state with $N$ fermions, each with filling fraction $\nu$.

It should be noted that the conductivities computed here are independent of the temperature.  The tip of the D8-brane is away from the horizon, and therefore the condition that the charge at the tip vanishes should not depend on the horizon's exact position.  In particular, in terms of the two-component description of the current carriers \cite{Lifschytz:2009si, Bergman:2010gm}, all of the charge is of the dissipationless type, smeared along the radial direction of the D8-brane.  The strings stretching from the horizon to the D8-brane are massive, and fluctuations producing charge-anticharge pairs are gapped.\footnote{This somewhat intuitive explanation of the mass gap has been verified in the D3-D7' model by a more careful fluctuation analysis \cite{Jokela:2010nu}.  We will leave the study of the fluctuations of the D2-D8' system for a future work.}

We have found a vanishing longitudinal conductivity.  However, this method only takes into account massless charge carriers.  The QH state is gapped, and the lightest charged excitations which could support a current are thermally pair-produced strings stretching from the horizon to the D8-brane.   Because such strings have masses of the order of $m_{gap}$, the Boltzmann factor for such fluctuations is of the form $e^{-C m_{gap}/T}$, for some unknown constant $C$.   Including such suppressed thermal pair production would likely yield $\sigma_{xx} \sim e^{-C m_{gap}/T}$.  Such an exponentially-suppressed conductivity is observed in real quantum Hall systems \cite{vonKlitzing, Tsui}.

\subsection{Metal}
\label{sec:BHconductivity}
For the BH embeddings, we can compute the longitudinal and Hall conductivities using the standard Karch-O'Bannon method \cite{Karch:2007pd, O'Bannon:2007in}; the currents are fixed by the condition that the action be real.  In particular, this requires that $X$ have double roots at the pseudohorizon $u_*$, which implies the following three conditions:
\bea
\label{firstcondition}
0 &=&  1+\frac{h^2}{u_*^5}-\frac{e^2}{f_*u_*^5}  \\
\label{secondcondition}
e\tilde j_{y*} &=& f_*h\tilde d_* \\
\label{thirdcondition}
 j_x^2 + \tilde j_{y*}^2 &=& f_*u_*^5\cos^6\psi_* \left(b^2u_* +\sin^4\psi_*\right) + f_*\tilde d_*^2  \ .
\eea

While these conditions, in principle, give the full nonlinear conductivity, we will specialize to the case of small electric field to obtain analytic expressions for the linear response.  The first condition (\ref{firstcondition}) gives the location of the pseudohorizon which, to leading order in $e$, is
\be
u_* \approx u_T \left(1+\frac{e^2}{u_T^2 + h^2}\right) \quad .
\ee
The thermal factor at the pseudohorizon, to leading order in $e$, is then
\be
f(u_*) \approx \frac{e^2}{u_T^5 + h^2} \ .
\ee
The second condition (\ref{secondcondition}) gives the physical, linear Hall conductivity
\be
\sigma_{xy} =  \frac{(2\pi\alpha')^2\mathcal N}{L^4 V_3}\frac{j_y}{e} =  \frac{3 N}{4\pi} \left(\frac{h}{u_T^5 + h^2} \tilde d + c(\psi_T) \right) \quad ,
\ee
and the third condition (\ref{thirdcondition}), when combined with the condition for the Hall current (\ref{secondcondition}), gives the physical, linear longitudinal conductivity
\be
\label{sigmaxxmetal}
\sigma_{xx} = \frac{(2\pi\alpha')^2\mathcal N}{L^4 V_3} \frac{j_x}{e} =  \frac{3 N}{4\pi}  \frac{u_T^{5/2}}{u_T^5 + h^2} \sqrt{\tilde d^2 + \cos^6\psi_T\left(b^2u_T +\sin^4\psi_T\right)\left(u_T^5 +h^2\right)} \ . 
\ee

\begin{figure}[ht]
\center
\includegraphics[width=0.48\textwidth]{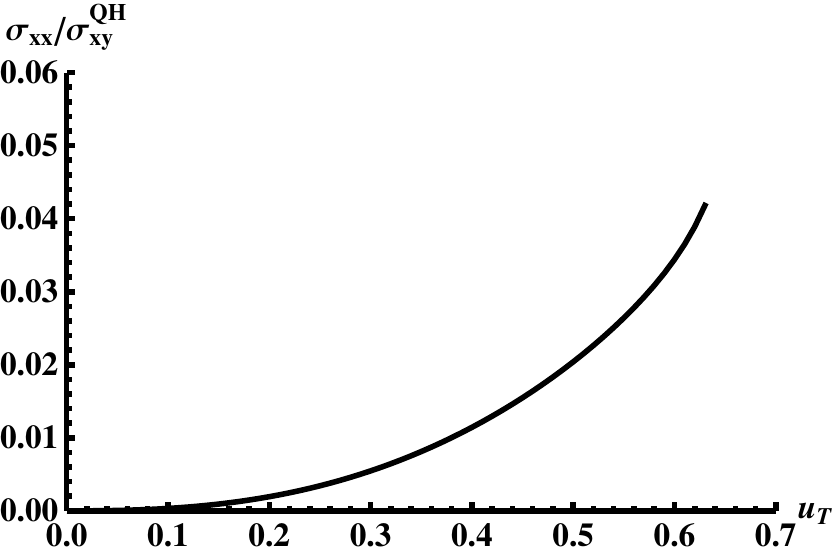}
\hspace{0.3cm}
\includegraphics[width=0.48\textwidth]{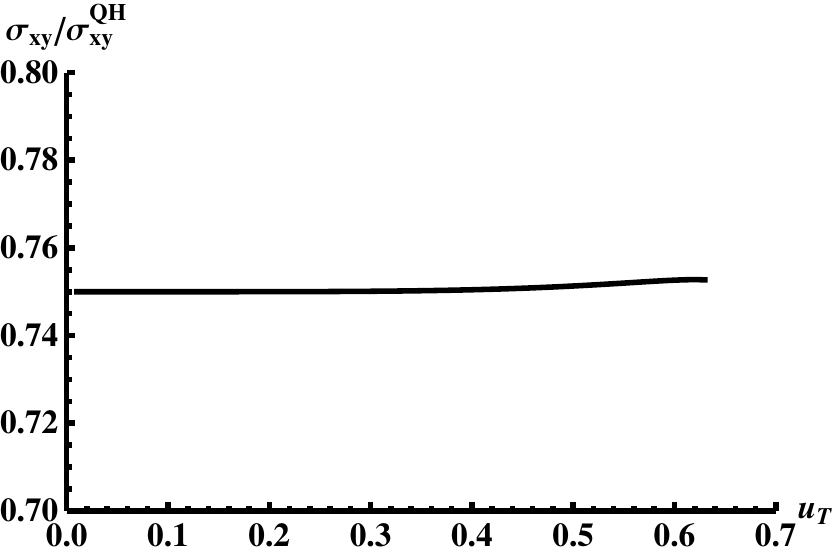} 
\caption{For the metallic phase, the conductivity is plotted as a function of temperature with $h=2$, $m= 0.1$, and $b=d=1$ in units of the Hall conductivity in the QH state $\sigma^{QH}_{xy}$.  On the left, the longitudinal conductivity $\sigma_{xx}$ scales as $u_T^{5/2}$ at low temperatures.  On the right, the Hall conductivity $\sigma_{xy}$ is approximately independent of temperature. }
\label{fig:metalconductivity}
\end{figure}

We numerically evaluate the conductivities and display the results in Fig.~\ref{fig:metalconductivity}.  Both the longitudinal and Hall conductivities are nonzero, as one would expect for a metallic phase.  Note that as $d/h \to 2/3$, the conductivities smoothly approach their QH values, (\ref{sigmaxxQH}) and (\ref{sigmaxyQH}).  As was seen in both the D3-D7' model \cite{Bergman:2010gm} and the Sakai-Sugimoto model \cite{Lifschytz:2009si}, the Hall current is made of contributions from two types of current carriers; one with charge density $\tilde d$ is located at the horizon and loses energy to the heat bath, while the rest of the charge density, $hc(\psi_T) = d-\tilde d$, is distributed along the D8-brane above the horizon and suffers no dissipation.

\begin{figure}[ht]
\center
\includegraphics[width=0.7\textwidth]{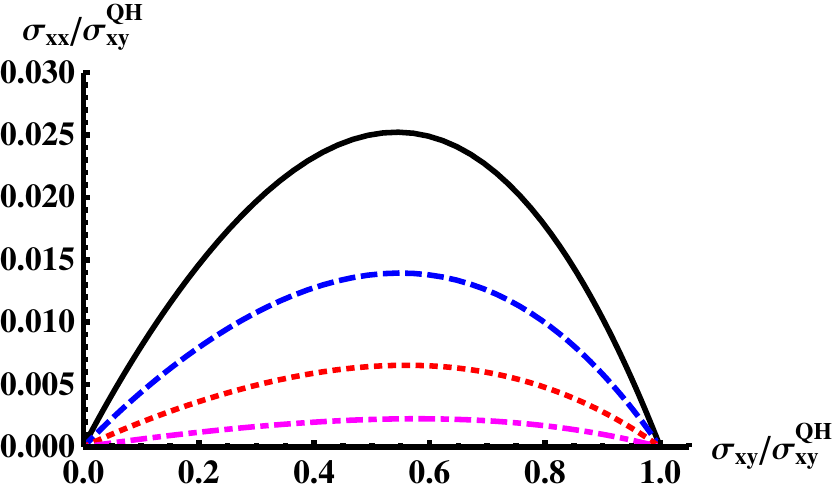}
\caption{The longitudinal conductivity $\sigma_{xx}$ is plotted against the Hall conductivity $\sigma_{xy}$ for varying magnetic field $h$ in units of the Hall conductivity of the QH state $\sigma^{QH}_{xy}$ at $b=d=1$ and $m=0.1$.  The temperatures are, from top to bottom, $u_T = 0.5$ (solid black), 0.4 (dashed blue), 0.3 (dotted red), and 0.2 (dot-dashed pink).  For $h=1.5$, at the MN embedding, $\left(\sigma_{xy}/\sigma^{QH}_{xy}, \sigma_{xx}/\sigma^{QH}_{xy} \right) = (1,0)$.   The curve proceeds to the left as $h$ increases, and for $h  \to \infty$, the curve approaches the origin.}
\label{fig:semicircle}
\end{figure}

One seemingly non-metallic property is the low-temperature behavior of the longitudinal conductivity.  From (\ref{sigmaxxmetal}) we see that, for vanishing magnetic field, $\sigma_{xx}$ diverges as $u_T^{-5/2}$.  This is due to the unbroken translational symmetry of the D2-D8' system; we have introduced neither impurities nor edges, both of which would break translational invariance.  Similarly, (\ref{sigmaxxmetal}) implies that when $h$ is nonzero, $\sigma_{xx}$ vanishes as $u_T^{5/2}$, which is a consequence of the fact that at $T=0$ the full Lorentz invariance is restored and the longitudinal conductivity must vanish.  For a real metal, $\sigma_{xx}$ approaches a finite, nonzero value at small temperatures\footnote{This assumes, of course, that there is no transition to a superconducting phase.} due to scattering off impurities.  Here, charge carriers are only scattered by thermal fluctuations and so behave at zero temperature as free particles.

In a quantum Hall system at low temperature, as the magnetic field is varied from one plateau to another, the conductivities sweep out a semicircle when plotted on the $(\sigma_{xy},  \sigma_{xx})$-plane \cite{Dykhne}.  The D2-D8' model contains only a single QH state and therefore strictly speaking can not model the transition between such states or reproduce this semicircle behavior.  However, in a system exhibiting only the IQHE, as the magnetic field is increased beyond the $\nu=1$ state, the system becomes a quantum Hall insulator and the conductivities trace a semicircle which approaches the origin, $\sigma_{xx} = \sigma_{xy}= 0$, at large magnetic field \cite{Hilke1, Hilke2}.  As illustrated in Fig.~\ref{fig:semicircle}, the D2-D8' model qualitatively reproduces this behavior.  One important discrepancy, apparent in (\ref{sigmaxxmetal}), is that, rather than reaching a maximum at $\sigma_{xx} = 1/2$, the maximum of the ``semicircle'' scales at low temperature as $u_T^{5/2}$.  We interpret this as a probable consequence of the unbroken translational invariance.

\section{Phase structure}
\label{sec:phasestructure}
Our first task in order to investigate the phase structure of the D2-D8' system is to compute the free energy.  The regulated, on-shell Euclidean bulk action gives the grand potential of the boundary theory:
\be
\Omega(\mu,T,h) = \frac{1}{\mathcal N} S^E_{on-shell} \nonumber\\
=\frac{1}{\mathcal N} \left(S^E_{DBI+CS}+ S^E_{bdry} + S_{CT}\right) \big|_{on-shell} \ ,
\ee
where $\mu$ is the chemical potential and $S_{bdry}$ is the boundary term given by (\ref{boundaryaction}).  We regulate the divergences of the action by adding a gauge and diffeomorphism invariant counterterm\footnote{Any such counterterm with these symmetries and which cancels the divergences will suffice.} 
\be
S_{CT} = - \frac{L T_8 g_s^{4/5}}{4}  \int d\Sigma \ e^{-\frac{4\phi}{5}} \sqrt{\det (\gamma_{\mu\nu} +2\pi\alpha' F_{\mu\nu})}
\ee 
which is integrated over the UV cutoff surface $\Sigma$ with induced metric $\gamma$.  By choosing the cutoff to be the surface $u = u_{UV}$, the counterterm is then
\be
S_{CT} = - \mathcal N|b|\frac{u_{UV}^4}{4}\left(1+2\psi_{UV}^2\right)
\ee
which cancels the divergences but also contributes a finite piece proportional to $|b|mc$.

Since we are working in the canonical ensemble, we need to perform a Legendre transform to obtain the free energy:
\be
F(d,T,h) = \Omega(\mu(d),T,h) + d\mu \ .
\ee
We can write the chemical potential $\mu(d) = a_0(\infty) = \int^\infty_{u_{min}} du \, a_0' + a_0(u_{min})$, such that
\bea
F &=& \int du \, \left(u^\frac{5}{2} \cos^3\psi \sqrt{\left(1+fu^2\psi'^2-a'^2_0\right)\left(b^2u + \sin^4\psi\right)\left(1 + \frac{h^2}{u^5}\right)}  \right.\nonumber \\
& & \hspace{0.5cm} + \, da_0' - c(\psi) ha_0' \Bigg) + \frac{1}{\mathcal N}S_{CT} + da_0(u_{min}) - c(\psi(u_{min})) ha_0(u_{min}) \ .
\eea
For BH embeddings, $a_0(u_T) = 0$, so the last two terms vanish.  For MN embeddings, recall that $\tilde d = d-c(\pi/2)h$, and so the last two terms cancel.  The free energy is then
\bea
F &=& \int du \, \left(u^\frac{5}{2} \cos^3\psi \sqrt{\left(1+fu^2\psi'^2-(a_0')^2\right)\left(b^2u + \sin^4\psi\right)\left(1 + \frac{h^2}{u^5}\right)} + \tilde da_0'  \right) \nonumber \\
&&  + \frac{1}{\mathcal N}S_{CT}  \ .
\eea
We can now compute the free energy for the various embeddings and, by comparing them, determine the thermodynamically preferred state as a function of the external parameters.

\subsection{High-temperature phase transition}
\label{sec:hightemptransition}
\begin{figure}[ht]
\center
\includegraphics[width=0.65\textwidth]{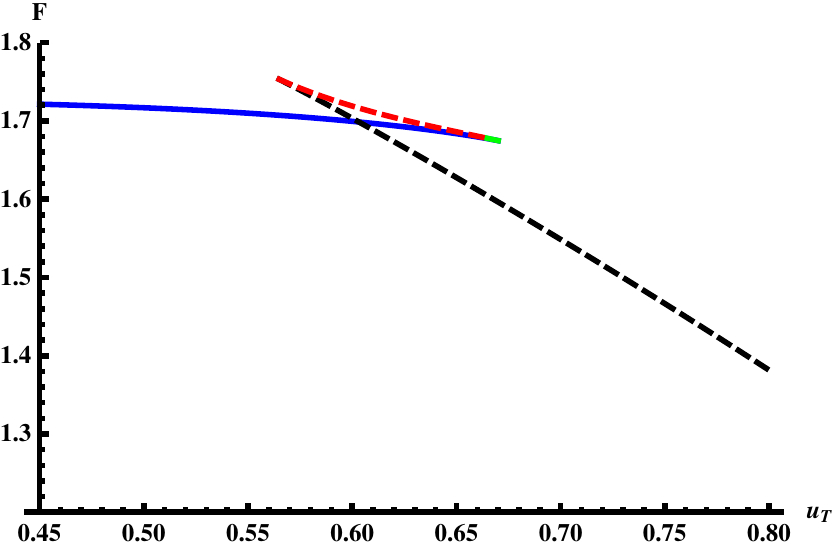} 
\caption{A plot of the free energy $F$ as a function of the temperature $u_T$ in the locked phase, $d/h=2/3$, for $m=0.1$ and $ b=d=1$.  The solid blue curve shows the stable MN embedding, the dashed black curve is the stable, small $\psi_T$ BH embedding, while unstable MN solution is solid green and the unstable BH embedding is dashed red.  The swallowtail structure indicates a first-order phase transition at $u_T\approx 0.6$.}
\label{fig:freeenergy}
\end{figure}

We begin with the special case where the magnetic field and charge are locked, $d/h=2/3$ and find a first-order phase transition from a MN embedding to a BH embedding when the temperature is increased.  This critical temperature is approximately the mass gap, $u_T^{crit} \lesssim u_0$.  The free energy as a function of temperature, shown in Fig.~\ref{fig:freeenergy}, takes the standard swallowtail form; near the phase transition, there exists an additional unstable phase which, depending on the temperature, is either a MN embedding at $u_0$ close to $u_T$ or a BH embedding with $\psi_T$ close to $\pi/2$.

\begin{figure}[ht]
\center
\includegraphics[width=0.65\textwidth]{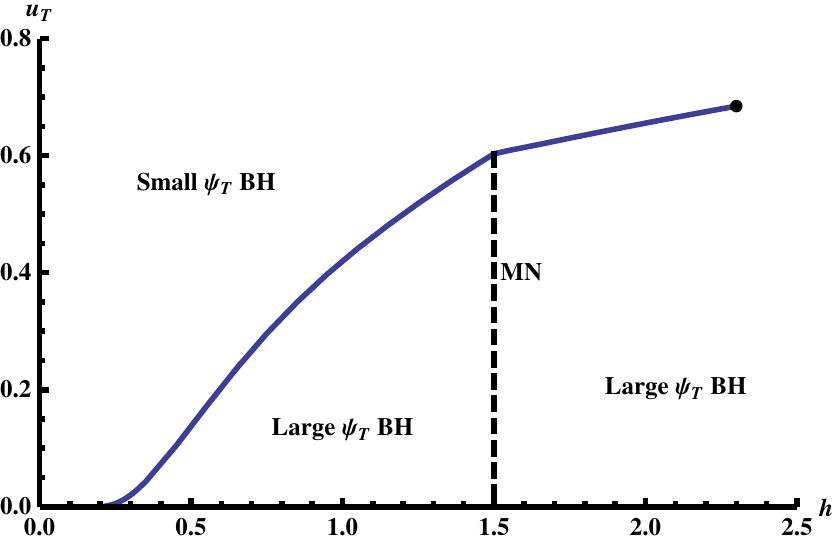} 
\caption{Phase diagram in the temperature-magnetic field plane for $b=d=1$ and $m=0.1$.  The solid blue curve is a line of first order-phase transitions from large $\psi_T$ spiky BH embeddings at low temperature to small $\psi_T$ embeddings at high-temperature.  This line ends in a second-order critical point $(h, u_T) \approx (2.3, 0.69)$  indicated by a black dot.  The dashed black vertical line indicates the locked phase, $d/h=2/3$ where there is a MN solution rather than a large $\psi_T$ BH embedding.}
\label{fig:phasediagram}
\end{figure}

A similar first-order phase transition was seen in the D3-D7' model \cite{Bergman:2010gm}.  In real quantum Hall fluids, however, when the temperature is of the order of the mass gap, the systems smoothly crosses over to metallic behavior \cite{vonKlitzing, Tsui}.


For generic charge and magnetic field, the low temperature phase is no longer a MN embedding but instead a large $\psi_T$ BH embedding. 
There is still a first-order phase transition as the temperature is increased, but now it is from a BH embedding with large $\psi_T$ to one with small $\psi_T$.  The phase diagram for the temperature-magnetic field plane is shown in Fig.~\ref{fig:phasediagram}.  For fixed $d$, the critical temperature grows with $h$.  The line of first-order phase transition ends at a critical point; for sufficiently large $h$, there is always only a single BH embedding, and $\psi_T$ decreases smoothly with increasing temperature.

\begin{figure}[ht]
\center
\includegraphics[width=0.48\textwidth]{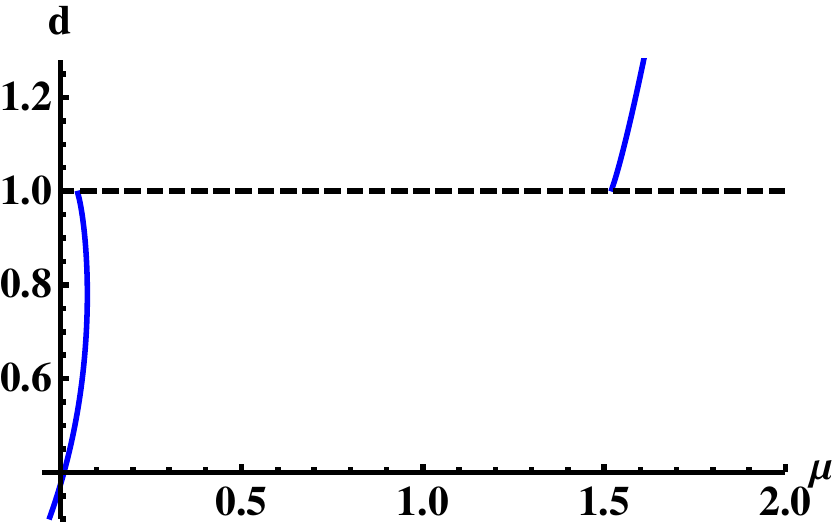}
\hspace{0.3cm}
\includegraphics[width=0.48\textwidth]{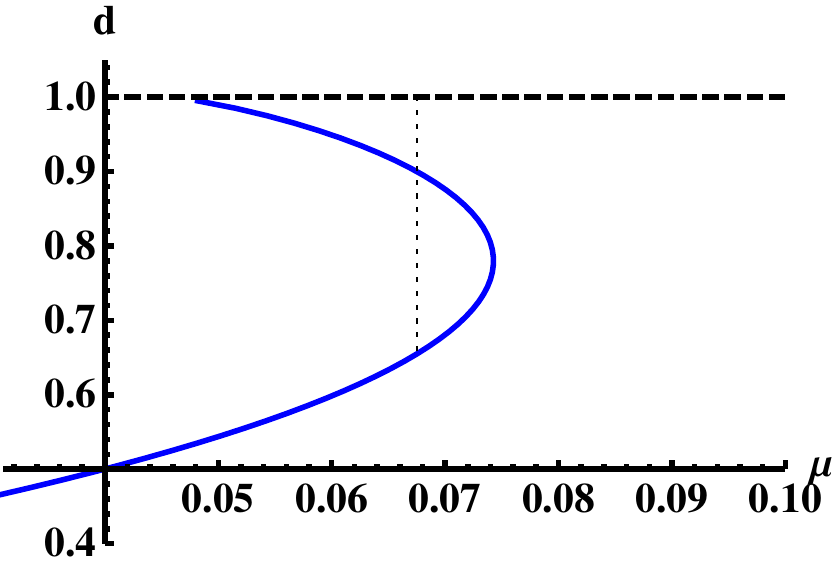}
\caption{The charge density $d$ as a function of the chemical potential $\mu$  for fixed $h = 3/2$, $b=1$, $u_T = 0.1$, and $m=0.1$.  Away from $d=1$, the embedding is a large $\psi_T$ BH (blue curve), while at $d=1$ it is a MN embedding with mass gap $u_0 \approx 0.83$ (dashed black line).  The range in $\mu$ without a BH solution is of size $\Delta \mu  \approx 1.46$.  The plot on the right zooms in to show that for $\mu$ between 0.048 and 0.074, there are two BH solutions in addition to the MN solution.  The dotted vertical line illustrates the Maxwell construction, by which we can determine that the first-order phase transition occurs at $\mu \approx 0.068$.}
\label{fig:muvsd}
\end{figure}

\subsection{Transition from the QH state}
\label{sec:transitionfromlocking}
There is also a transition from the QH state to a metallic state, that is from a MN solution to the large $\psi_T$ BH embedding, as either the charge or magnetic field is varied away from $d/h = 2/3$.  A similar transition from MN to BH phases occurs in the supersymmetric D3-D7 model, where the MN embedding only exists at zero density \cite{Ghoroku:2007re, Mateos:2007vc}.

The free energy has a kink as the system passes through the MN solution; at fixed the magnetic field, the chemical potential $\mu = \frac{\partial F}{\partial d}$ appears to be discontinuous across $d = \frac{2}{3} h$, as shown in Fig. \ref{fig:muvsd}.   The jump in $\mu$ can be calculated by noticing that for very spiky BH solutions, $a_0' =  \pm 1$ along the length of the spike, with the sign depending on whether $d/h \to 2/3$ from above or below.  Away from the spike, the limit of $a_0'$ as $d/h \to 2/3$ is otherwise smooth.  Since the length of the spike is $u_0 - u_T$, we find $\Delta \mu = 2(u_0 - u_T)$.  

This discontinuity is an artifact of working in the canonical ensemble.  The problem is that the MN solution exists at a single $d$ but for any $\mu$.  Because the D7-brane avoids the horizon, $a_0$ does not have a Dirichlet IR boundary condition, and $a_0(\infty) = \mu$ is unfixed by the dynamics and can take any value.  As a result, the canonical ensemble is not well defined.

We will switch instead to the grand canonical ensemble, where the nature of the transition can be more clearly understood.  MN solutions exist for all $\mu$, but there is a range in $\mu$ of size $\Delta \mu$ where there are no BH embeddings.\footnote{Note that $\Delta \mu = 2(u_0 - u_T)$ is equal to the mass of a quark plus the mass of an antiquark.}  In this particular range, the system can only be in the MN phase.   But, where the two solutions coexist, the BH phase has lower grand potential $\Omega$ and is therefore preferred.  At the large-$\mu$ edge of this range, $d$ is linearly increasing with $\mu$ and so is continuous as the system goes from the MN solution to the BH solution; the phase transition is therefore marginally second-order.  However, at the small-$\mu$ end, as can be seen in the zoomed in plot in Fig. \ref{fig:muvsd}, there are two BH solutions for a narrow range of $\mu$.   The branch which connects to the MN solution has $\frac{\partial d}{\partial \mu} < 0$, meaning it is unstable.\footnote{In an upcoming work on the fluctuation spectrum of this model, we will investigate the nature of this instability.}  As $\mu$ is lowered, the system jumps from the MN solution to the stable BH solution at lower $d$, thereby undergoing a first-order phase transition.  The critical $\mu$ at which the transition occurs can be found via a Maxwell construction, as illustrated in Fig. \ref{fig:muvsd}.

There is a surprising asymmetry between the transition away from the QH phase toward larger $d$ and to smaller $d$.  In the supersymmetric D3-D7 model \cite{Ghoroku:2007re, Mateos:2007vc}, where the analogous transition occurs at $d=0$, the transition   could be either first- or second-order depending on the temperature.  However, at a given temperature the transition to positive and negative $d$ was always identical due to charge conjugation symmetry.  Here, this symmetry has been broken by our choice of $h$, although it remains unclear precisely why these transitions should be of different orders.

\section{Discussion}
\label{sec:discussion}
We have presented a holographic model of a 2+1-dimensional, strongly-coupled fermion system at nonzero temperature, magnetic field, and charge density which exhibits quantum Hall behavior with integer filling fraction.  The gapped QH states of this D2-D8' system correspond to MN embeddings of the D8-brane, which do not enter the horizon.  We found in Section \ref{sec:MNembeddings} that these solutions exist only at low temperature, $u_T \lesssim m_{gap}$, and for a fixed ratio of the charge density to the magnetic field, $d/h = 2/3$, which implies $\nu = 1$.   The mass gap scales with the magnetic field as $h^{0.36}$.   The Hall conductivity of this state, computed in Section \ref{sec:QHconductivity}, takes the QH form $\sigma_{xy} = \frac{N\nu}{2\pi}$, while the longitudinal conductivity vanishes.  

At higher temperatures and at generic $d/h$, the D8-brane is in a BH embedding which corresponds to a metallic phase, similar to that found in the transition between two QH plateaux.  In Section \ref{sec:BHconductivity}, the conductivity was found to somewhat reproduce the semicirlce behavior seen in QH transitions.  However, the D2-D8' system shows unphysical behavior at the low temperature due to unbroken translation invariance.

In Section \ref{sec:hightemptransition}, we found a first-order phase transition from the QH state to the metallic state with a critical temperature of the order of the mass gap.  This phase transition persists away from the MN solution, for generic charges and magnetic fields, as a transition between two different BH embeddings.  At large enough magnetic field, the line of first-order phase transitions ends in a critical point.  

When the charge or magnetic field is varied away from the MN solution, there is a phase transition, as seen in Section \ref{sec:transitionfromlocking}.  In the grand canonical ensemble, the transition away from the QH phase is first-order when going to smaller $\mu$ and second-order when going to larger $\mu$.

Having constructed a QH state with $\nu = 1$, one might inquire whether this model might include other integer filling fractions as well.   As we discussed in Section \ref{sec:MNembeddings}, for a MN embedding all the charge must be induced via the Chern-Simons term, which is in turn directly related to how much of the $S^2\times S^3$ is wrapped by the D8-brane.  This quantity is topological in that it only depends on the endpoints of the D8-brane embedding; since the tip must be at $\psi = \pi/2$ and the UV asymptotics demand $\psi_\infty = 0$, the entire internal manifold is wrapped and $\nu = 1$.\footnote{This is in contrast to the D3-D7' model where $\psi_\infty$ can take a range of values, leading to filling fractions between approximately 0.7 and 0.8.}

Hypothetically, one could obtain higher-integer filling fractions from multiply-wrapped embeddings.  Such embeddings, like the negative mass embeddings discussed in Section \ref{sec:embeddings}, would self-intersect and suffer conical singularities when $\psi$ passes through zero.  At these points, the low-energy description of the D8-brane would break down as a result of tachyons, and the embedding would most likely be unstable.

\bigskip
\noindent

{\bf \large Acknowledgments}

We would like to  thank Oren Bergman, Brian Dolan, Sung-Sik Lee, and Gilad Lifschytz for helpful discussions, comments, and suggestions.
N.J. is supported in part by the Israel Science Foundation under grant no.~392/09 and in part at the Technion by a fellowship from the Lady Davis Foundation. 
M.J. is supported in part by Regional Potential program of the E.U.FP7-REGPOT-2008-1: CreteHEPCosmo-228644 and by Marie Curie contract PIRG06-GA-2009-256487.
The research of M.L. is supported by the European Union grant FP7-REGPOT-2008-1-CreteHEPCosmo-228644 and in part by the APCTP Focus Program Aspects of Holography and Gauge/string duality at the Asia Pacific Center for Theoretical Physics (APCTP).  M.L. would like to thank the APCTP and the Galileo Galilei Institute for Theoretical Physics for their hospitality while this research was in progress.

\appendix

\section{UV asymptotic analysis}
\label{UVasymptotics}
We investigate asymptotic solutions for the  D8-brane embedding $\psi(u)$ at large $u$.  Considering first the case without internal flux $b=0$ and taking the ansatz $\psi \sim \psi_\infty + Au^{\Delta}$ where we assume $\mathrm{Re}\, \Delta < 0$, to leading order $\mathcal O\left(u^\frac{5}{2}\right)$, the equation of motion (\ref{psieom}) is
\be
0 = -3 \cos^2\psi_\infty\sin^3\psi_\infty + 2\cos^4\psi_\infty\sin\psi_\infty
\ee
which implies three possible solutions for $\psi_\infty$
\be
\psi_\infty =  0, \pi/2, \arctan\left(\sqrt{\frac{2}{3}}\right) \quad .
\ee
We can immediately exclude the trivial constant solutions $\psi = 0$ or $\pi/2$, because those correspond to D8-branes with vanishing volume and energy. 

The equation of motion to order $\mathcal O(u^{5/2+\Delta} )$ gives
\bea
\label{firstorderUVeomzerob}
-\Delta(\Delta+7/2)\sin^2\psi_\infty \cos^3\psi_\infty &=& 3  \sin^3\psi_\infty \cos^2\psi_\infty(2\tan\psi_\infty-3\cot\psi_\infty)  \nonumber \\
&&-2 \sin\psi_\infty \cos^4\psi_\infty(4\tan\psi_\infty-\cot\psi_\infty)\quad .
\eea
For $\psi_\infty = 0$, this equation cannot be satisfied, so this is not, in fact, a solution.  At this order $\psi_\infty = \pi/2$ is still a solution, but it similarly fails at order $\mathcal O(u^{5/2+2\Delta})$.  Furthermore, it can be shown that any solution with $\psi_\infty = 0$ or $\pi/2$ must be constant and therefore degenerate.  

For the third solution $\psi_ \infty = \arctan\left(\sqrt{\frac{2}{3}}\right)$, equation (\ref{firstorderUVeomzerob}) reduces to
\be
\Delta^2+\frac{7}{2}\Delta+10=0 \quad ,
\ee
which has the complex solutions
\be
\Delta = -\frac{7}{4} \pm i \frac{\sqrt{111}}{4} \quad .
\ee
Thus, the $b=0$ embedding is unstable as expected.

To stabilize the embedding, we now consider $b>0$.   The leading term is linear in $b$, meaning that (\ref{psieom}) is asymptotically
\be
0 = -3 b u^3 \cos^2\psi_\infty\sin\psi_\infty  \quad , 
\ee
and the only solutions are $\psi_\infty = 0$ and $\pi/2$.

First, let us look at  the $\psi_\infty = 0$ case.  To the next order, the equation of motion (\ref{psieom}) is then:
\be
Ab\Delta(4+\Delta)u^{3+\Delta} = -3 Ab u^{3+\Delta} + 2 A^3 b^{-1} u^{2+3\Delta} \quad .
\ee
Because we have assumed $\mathrm{Re}\, \Delta<0$, the second term on the RHS is subleading, so we find
\be
 \Delta^2 + 4\Delta + 3 = 0 \quad ,
 \ee
which implies $\Delta = -1$ or $-3$.  Note that $\Delta$ is real, as demanded by stability, for all nonzero $b$.

Now, we will consider the $\psi_\infty = \pi/2$ solution.  The asymptotic equation of motion (\ref{psieom}) is then:
\be
-A^4b\Delta(4+4\Delta)u^{3+4\Delta} = -3 A^2b u^{3+2\Delta} + 2 A^4 b^{-1} u^{2+4\Delta} \quad .
\ee
Again, the second term on the RHS is subleading, so we obtain
\be
-A^2\Delta(4+4\Delta) = 3u^{-2\Delta} \quad ,
\ee
which has not solutions for $\mathrm{Re}\, \Delta<0$; thus, the power law ansatz is not a valid solution for $\psi_\infty =\pi/2$.  In fact, it can be shown that any nontrivial embedding will fail to satisfy the equation of motion, and therefore, only the degenerate constant solution $\psi = \pi/2$ has this asymptotic behavior.

\end{document}